\documentclass[%
 reprint,
 showpacs,
 preprintnumbers,
 amsmath,amssymb,
 aps,
prl,
]{revtex4-1}
\usepackage{amsmath}
\usepackage{graphicx}
\usepackage{dcolumn}
\usepackage{bm}
\usepackage{color}
\usepackage{multirow}
\usepackage{hyperref}

\begin{document}


\title{Discrete Transformation of Baryon- and Lepton-Nonconserving Processes}

\author{Ehsan Jafari}
\email{ehsan.jafari@uky.edu}
\affiliation{Department of Physics and Astronomy, University of Kentucky,
Lexington, KY 40506, USA}

\date{\today}

\begin{abstract}

We consider discrete transformations  ($C, P, T$ and $CP$) of baryon- and lepton-nonconserving processes.
It has long been thought that values ($\pm 1, \pm i$) form the correct set of fermionic arbitrary phases for discrete
transformations. In this paper we show that this idea is not generally true. In order to count the number of fundamental fermions in a
process, $F$ number has been introduced. According to our evaluation for any operator which breaks $B-L$ symmetry and violate $F$
number the set of values ($\pm 1, \pm i$) is not the correct set of fermionic arbitrary phases,
due to the fact that discrete transformations of such operators will change by altering the fermionic phases from $\pm 1$ to $\pm i$.

\end{abstract}
\pacs{Valid PACS appear here}
\maketitle
\section{Introduction}
Although baryon number ($B$) and lepton number ($L$) are broken in the standard model ($SM$) but the difference between them,
$B-L$, is a conserved quantum number in this model. If we consider beyond the $SM$ scenarios, $B-L$ symmetry would be broken as well.
One of the outcomes of grand unified theories is baryon and lepton non-conserving interactions.
Operator analysis of baryon- and lepton-nonconserving processes has been done for the first  time by
Weinberg \cite{weinberg1979baryon} and Wilczek and Zee \cite{wilczek1979operator}. A detailed generalization of these operators
has been considered for different possible processes in the references \cite{weinberg1980varieties,weldon1980operator}. Transition
of neutron to anti-neutron has been studied with more details in the references \cite{kuo1980neutron,rao1982n,caswell1983matter}.

In this paper we collect baryon- and lepton-nonconserving operators from different references and evaluate discrete
transformations; charge conjugation ($C$), parity ($P$), time reversal ($T$) and charge conjugation-parity
$CP$; of these operators. Study of $CP$ transformation enables us to understand if the basic laws of physics
are the same for matter and anti-matter. $C$ and $CP$ transformation of baryon- and lepton-nonconserving processes is also matter
of interest for any scenario of baryogenesis and leptogenesis. To start the universe from a symmetric state between matter and
anti-matter and evolve it to the current completely asymmetric state, we need interactions which violate baryon and lepton numbers as
well as $C$ and $CP$ symmetries \cite{sakharov1967violation}.

To evaluate the number of quarks and leptons as the constituents of particles in interaction, we define fundamental
fermion number ($F$ number). Each of the quarks and leptons of any generation are assigned $F=1$ and each of the anti-quarks and
anti-leptons of any generation are assigned $F=-1$. Thus, $F$ number is related to baryon number ($B$) and lepton number ($L$) by: $F=3B+L$.
For instance proton has $F(p)=3$, positron has $F(e^+)=-1$ and pions of any charge have $F(\pi^{+,0,-})=0$. In the same way that we
evaluate $B$ and $L$ conserving or violating interactions, we can consider $F$ conserving or violating interactions.

As we know, if we have the correct sets of fermionic and bosonic  arbitrary phases for discrete transformations, choosing
any values of these sets as fermionic and bosonic phases, should leads to the same discrete transformation. It has long been thought that the values ($\pm 1, \pm i$)
form the correct set of fermionic arbitrary phases for discrete transformations of any Hamiltonian.
According to our calculations this old believe is not true. For the processes which break $B-L$ symmetry and violate $F$ number,
switching fermionic phases from $\pm1$ to $\pm i$ will change discrete transformations of operators as well.
Therefore, for these kind of interactions, the set of values ($\pm 1, \pm i$) do not form the correct set of
fermionic arbitrary phases.

As we know $SM$ fails to explain the amount of observable asymmetry between matter and anti-matter in the universe \cite{bernreuther2002cp}
and we should use beyond the $SM$ theories to explain this asymmetry. In the beyond the $SM$ theories, $B-L$ is not a
conserve quantum number. Based on our results, for the baryon- and lepton-nonconserving processes which breaks $B-L$ symmetry
and violates $F$ number, the set ($\pm 1, \pm i$) is not the correct set of fermionic arbitrary phases.

Recently discrete transformation of neutron oscillation was a matter of interest for experts in this field \cite{berezhiani2015neutron,fujikawa2015neutron, mckeen2015cp}.
Neutron oscillation is an example of the the processes which breaks $B-L$ symmetry and violate $F$ number. In these papers $n-\bar{n}$ oscillation
has been studied in the low energy limit and neutron and antineutron has been considered as a single particle. Berezhiani and Vainshtein
have claimed that CP violation is necessary for neutron oscillation process \cite{berezhiani2015neutron}. This claim has been rejected by Fujikawa
and Tureanu \cite{fujikawa2015neutron} which followed by another paper by McKeen and Nelson \cite{mckeen2015cp}.
In the references \cite{fujikawa2015neutron,mckeen2015cp} it is proved that we can not determine if neutron oscillation is $CP$ violating or
$CP$ conserving.

\section{Fermionic Field Operator And Discrete Transformations}
If we show fermionic field operator by $\psi(t,x)$ and consider the metric $g^{\mu\nu}=(1,-1,-1,-1)$ and $\gamma$ matrices with
Dirac algebra $\left\{\gamma^{\mu},\gamma^{\nu}\right\}=2g^{\mu\nu}$, we can define transformation of $\psi(t,x)$ under the action
of discrete transformations as:
\begin{equation}
C\psi \left (t,x \right )C^{\dagger}=\eta_{c} c \bar{\psi}^{T} \left (t,x \right )
\end{equation}
\begin{equation}
P\psi \left (t,x \right )P^{\dagger}=\eta_{p} \gamma^{0} \psi \left (t,-x \right )
\end{equation}
\begin{equation}
T\psi \left (t,x \right )T^{-1}=\eta_t \gamma^{1}\gamma^{3} \psi\left (-t,x \right )
\end{equation}
where $c=i\gamma^2\gamma^0$ and $\eta_c$, $\eta_p$ and $\eta_t$ are arbitrary phases with the condition $|\eta_p|^2,|\eta_c|^2,|\eta_t|^2=1$.
These arbitrary phases can be narrowed down to $\eta^2_c, \eta^2_p, \eta^2_t =\pm1$ due to the fact that we expect applying two successive discrete
transformations should not change the observable quantities which all consist of even number of field operators \cite{peskin1995introduction}.
This leads to $\eta_c,\eta_p,\eta_t=\pm 1$ or $\pm i$.

It has long been assumed that the values ($\pm 1, \pm i$) form the correct set of fermionic arbitrary phases.
Therefore, we can see different choices of these values as fermionic arbitrary phases in literatures. As an example if we
consider $\bar\psi\Gamma\psi$ where $\Gamma=1,\gamma^5,\gamma^\mu,\gamma^\mu\gamma^5,\sigma^{\mu\nu}$; discrete
transformations of these fermion bilinears are the same if we choose any value of the set ($\pm 1, \pm i$) as an arbitrary phase.
In this paper we will show that this assumption is not generally true. We choose Peskin and Schroeder convention and put
$\eta_c=\eta_p=\eta_t=1$ \cite{peskin1995introduction}. We will see that changing these arbitrary phases from $1$ to $\pm i$ will
change discrete transformations of interactions which break $B-L$ symmetry and violate $F$ number. For the mentioned interactions if
we change any of the fermionic arbitrary phases $\eta_c$, $\eta_p$ or $\eta_t$ from $1$ to $\pm i$, terms which are $-even$ (or $-odd$)
under the action of discrete transformations become $-odd$ (or$-even$). Thus, the set ($\pm 1, \pm i$) is not the
correct set of fermionic arbitrary phases.

In this paper we have collected baryon- and lepton-nonconserving operators from the various references. We should emphasize that all of
these operators are model independent and in writing them  no specific grand unified theory or other baryon- and lepton-nonconserving gauge
theory has been considered. Table of discrete transformation of different fermion bilinears is provided in the Appendix. These fermion bilinears
are used as the building blocks of interacting operators.

\subsection{$\Delta B=\Delta L=-1$ nucleon decay($p\rightarrow e^{+}\pi^{0}, etc.$) \cite{weinberg1979baryon}}

These interactions violate $F$ number and conserve $B-L$ symmetry. Because of conservation of $B-L$ nucleon can only decay to anti-leptons,
$n\rightarrow e^+\pi^-$. The dominant $SU(3)\times SU(2)\times U(1)$ invariant operators which mediate such interactions are dimension
$6$ ($d=6$) operators and have the form $QQQL$:

\begin{equation}\label{4}
  \epsilon_{\alpha\beta\gamma}\epsilon_{ij}\left(d^{T}_{R\alpha} c u_{R\beta}\right)\left(q^{T}_{L\gamma i}c l_{Lj}\right),
\end{equation}
\begin{equation}
  \epsilon_{\alpha\beta\gamma}\epsilon_{ij}\left(q^{T}_{L\alpha i } c q_{L\beta j}\right)\left(u^{T}_{R\gamma}c e_{Rd}\right),
\end{equation}
\begin{equation}
  \epsilon_{\alpha\beta\gamma}\epsilon_{ij}\epsilon_{kl}\left(q^{T}_{L\alpha i} c q_{L\beta j}\right)\left(q^{T}_{L\gamma k}c l_{Ll}\right),
\end{equation}
\begin{equation}
\begin{split}
  \epsilon_{\alpha\beta\gamma}\left(q^{T}_{L\alpha i} c q_{L\beta j}\right)\left(q^{T}_{L\gamma k}c l_{Ll}\right)\\
           &\times \left(\vec{\tau}\epsilon\right)_{ij} . \left(\vec{\tau}\epsilon\right)_{kl},
\end{split}
\end{equation}
\begin{equation}
\epsilon_{\alpha \beta \gamma}\left(d^{T}_{R\alpha}cu_{R\beta}\right)\left(u^{T}_{R\gamma}ce_{R}\right),
\end{equation}
\begin{equation}\label{9}
  \epsilon_{\alpha \beta\gamma}\left(u^{T}_{R\alpha}cu_{R\beta}\right)\left(d^{T}_{R\gamma}ce_{R}\right).
\end{equation}

In the above equations $\alpha$, $\beta$ and $\gamma$ are $SU\left(3\right)$ indices, $i$, $j$, $k$ and $l$ are $SU\left(2\right)$
indices, $q_{L\alpha i}$ is the generic left-handed quark doublet, $u_{R\alpha}$ and $d_{R\alpha}$ are generic right-handed quark singlets with
charges $\frac{2}{3}$ and $-\frac{1}{3}$, $l_{Li}$ is the generic left-handed lepton doublet, $e_{R}$ is the generic right-handed charged lepton singlet,
 $\epsilon_{\alpha \beta \gamma}$ and $\epsilon_{ij}$ are totally antisymmetric $SU\left(3\right)$ and $SU\left(2\right)$ tensors with
 $\epsilon_{123}=\epsilon_{12}=1$.

The operators \eqref{4}-\eqref{9} have the same transformations under the action of $C$, $P$, $T$ and $CP$.
Each operator consists of four terms. According to the table of discrete transformations in the Appendix we can see that
terms with even or odd number of $\gamma^5$ have different transformations under the action of $C$ and $P$.
In the case of charge conjugation we have:

 \[
   C=\left \{
   \begin{tabular}{ccc}
   terms with even number of $\gamma^{5}$ are $C-even$\\
   terms with odd number of $\gamma^{5}$ are $C-odd$
   \end{tabular}
   \right.
 \]
 If instead of $\eta_c=1$ we choose $\eta_{c}=\pm i$ this result will not change. Also parity evaluation leads to:
 \[
   P=\left \{
   \begin{tabular}{ccc}
   terms with even number of $\gamma^{5}$ are $P-even$\\
   terms with odd number of $\gamma^{5}$ are $P-odd$
   \end{tabular}
   \right.
 \]
Adopting $\eta_{p}=\pm i$ will not  change this   transformations.

The above operators are $T-even$. This result would not be effected by changing $\eta_t$ from 1 to $\pm i$. Also, they are all $CP-even$.

As we can see, discrete transformations of these operators which conserve $B-L$ symmetry and violate $F$ number are
the same if we choose any value of the set ($\pm 1, \pm i$) as arbitrary phase. With the use of Appendix, we can see that under the action of discrete
transformations they acquire fermionic arbitrary phase coefficients $\eta_c^4$, $\eta_p^4$ and $\eta_t^4$ which is equal to $1$ for $\eta_c$,
$\eta_p$, $\eta_t=\pm1 $ or $\pm i$.

\subsection{$\Delta B=-\Delta L=-1$ nucleon decay($n\rightarrow e^{-}\pi^{+}$, etc.) \cite{weinberg1980varieties}}

These kind of processes violate $F$ number and break $B-L$ symmetry so, nucleons can decay to leptons($p\rightarrow e^-\pi^+\pi^+$).
To be invariant under Lorentz transformation and weak SU(2), the lowest dimensional operators for such interactions should have the
form $QQQ\Bar{L}B$ with $d=7$. Here $B$ is a boson field or spacetime derivative. The lowest dimensional $SU(3)\times SU(2)\times U(1)$
invariant operators are:

\begin{equation}\label{10}
\epsilon_{\alpha \beta \gamma} \epsilon_{ij} \epsilon_{kl} \left(q^{T}_{L\alpha i}cq_{L \beta j}\right)\left(\bar{l}_{Lk}d_{R \gamma}\right)\phi_{l}^{\dagger},
\end{equation}
\begin{equation}
\epsilon_{\alpha \beta \gamma} \left(q^{T}_{L\alpha i}cq_{L \beta j}\right)\left(\bar{l}_{Lj}d_{R \gamma}\right)\phi_{i}^{\dagger},
\end{equation}
\begin{equation}
\epsilon_{\alpha \beta \gamma} \left(d^{T}_{R\alpha}cd_{R \beta}\right)\left(\bar{e}_{R}q_{L\gamma i}\right)\phi_{i}^{\dagger},
\end{equation}
\begin{equation}
\epsilon_{\alpha \beta \gamma}\epsilon_{ij} \left(d^{T}_{R\alpha}cd_{R \beta}\right)\left(\bar{l}_{Li}u_{R\gamma}\right)\phi_{j}^
{\dagger},
\end{equation}
\begin{equation}
\epsilon_{\alpha \beta \gamma}\epsilon_{ij} \left(d^{T}_{R\alpha}cu_{R \gamma}\right)\left(\bar{l}_{Li}d_{R\beta}\right)\phi_{j}^
{\dagger},
\end{equation}
\begin{equation}\label{15}
\epsilon_{\alpha\beta\gamma}\left(d^{T}_{R\alpha}cd_{R\beta}\right)\left(\bar{l}_{Li}d_{R\gamma}\right)\phi_{i},
\end{equation}
\begin{equation}\label{16}
\epsilon_{\alpha\beta\gamma}\left(d^{T}_{R\alpha}cD_{\mu}d_{R\beta}\right)\left(\bar{l}_{Li}\gamma^{\mu}q_{L\gamma i}\right),
\end{equation}
\begin{equation}\label{17}
\epsilon_{\alpha\beta\gamma}\left(d^{T}_{R\alpha}cD_{\mu}d_{R\beta}\right)\left(\bar{e}_{R}\gamma^{\mu}d_{R\gamma}\right).
\end{equation}
\begin{equation}\label{18}
\epsilon_{\alpha\beta\gamma}\left(\bar{l}_{Li}D_{\mu}d_{R\beta}\right)\left(d^{T}_{R\alpha}c\gamma^{\mu}q_{L\gamma i}\right),
\end{equation}

Operators \eqref{10}-\eqref{18} do not have identical transformations under the action of discrete symmetries. Therefore we divide them into
three subgroups:

\subsubsection{Operators \eqref{10}-\eqref{15}}
Under the action of charge  conjugation we have:
\[
  C=\left \{
  \begin{tabular}{ccc}
  terms with even number of $\gamma^{5}$ are $C-even$\\
  terms with odd number of $\gamma^{5}$ are $C-odd$
  \end{tabular}
  \right.
\]

 If instead of $\eta_c=1$ we choose $\eta_{c}=\pm i$ then:
\[
  C=\left \{
  \begin{tabular}{ccc}
  terms with even number of $\gamma^{5}$ are $C-odd$\\
  terms with odd number of $\gamma^{5}$ are $C-even$
  \end{tabular}
  \right.
\]
As we can see due to changing $\eta_c$ from 1 to $\pm i$, $C-even$ terms became $C-odd$ and $C-odd$ terms became $C-even$. Under
the effect of parity we have:
\[
  P=\left \{
  \begin{tabular}{ccc}
  terms with even number of $\gamma^{5}$ are $P-odd$\\
  terms with odd number of $\gamma^{5}$ are $P-even$
  \end{tabular}
  \right.
\]
If we change $\eta_{p}$ from $1$ to $\pm i$, parity transformation will change as well:
\[
  P=\left \{
  \begin{tabular}{ccc}
  terms with even number of $\gamma^{5}$ are $P-even$\\
  terms with odd number of $\gamma^{5}$ are $P-odd$
  \end{tabular}
  \right.
\]
Because of changing $\eta_p$ from $1$ to $\pm i$, $P-odd$ terms become $P-even$ and vice versa.

This subgroup of operators are $T-even$ and if we choose $\eta_t=\pm i$ they become $T-odd$. According to our original phase
convention, they are $CP-odd$.

\subsubsection{Operators \eqref{16}-\eqref{17}}
Under the effect of charge conjugation we have:
\[
  C=\left \{
  \begin{tabular}{ccc}
  terms with even number of $\gamma^{5}$ are $C-even$\\
  terms with odd number of $\gamma^{5}$ are $C-odd$
  \end{tabular}
  \right.
\]
and if we change $\eta_c$ from $1$ to $\pm i$ we have:
\[
  C=\left \{
  \begin{tabular}{ccc}
  terms with even number of $\gamma^{5}$ are $C-odd$\\
  terms with odd number of $\gamma^{5}$ are $C-even$
  \end{tabular}
  \right.
\]
For parity we have:
\[
  P=\left \{
  \begin{tabular}{ccc}
  terms with even number of $\gamma^{5}$ are $P-odd$\\
  terms with odd number of $\gamma^{5}$ are $P-even$
  \end{tabular}
  \right.
\]
due to the change of $\eta_p$ from $1$ to $\pm i$ we have:
\[
  P=\left \{
  \begin{tabular}{ccc}
  terms with even number of $\gamma^{5}$ are $P-even$\\
  terms with odd number of $\gamma^{5}$ are $P-odd$
  \end{tabular}
  \right.
\]
Both of these operators are $T-odd$ and by changing $\eta_t$ to $\pm i$ they become $T-even$. Based on to our original arbitrary phase
convention these two operators are $CP-odd$.

\subsubsection{Operator \eqref{18}}
Under the effect of charge conjugation we have:
\[
  C=\left \{
  \begin{tabular}{ccc}
  terms with even number of $\gamma^{5}$ are $C-odd$\\
  terms with odd number of $\gamma^{5}$ are $C-even$
  \end{tabular}
  \right.
\]
by changing $\eta_c$ to $\pm i$, $C-odd$ terms become $C-even$ and vice versa. Under the effect of parity we have:
\[
  P=\left \{
  \begin{tabular}{ccc}
  terms with even number of $\gamma^{5}$ are $P-odd$\\
  terms with odd number of $\gamma^{5}$ are $P-even$
  \end{tabular}
  \right.
\]
and by changing $\eta_p$ to $\pm i$, $P-odd$ terms become $P-even$ and vice versa.

This operator is $T-odd$ and by choosing $\eta_t=\pm i$ it becomes $T-even$. By using our original phase convention this operator is
$CP-even$.

The operators \eqref{10}-\eqref{18} have the ability to destroy three quarks in the initial state and create one lepton
in the final state. On the other hand, they all break $B-L$ symmetry and violate $F$ number. As we have seen, for these operators
the values ($\pm 1, \pm i$) do not form the correct set of arbitrary phases due to the fact that different values of
this set will lead to the different transformations under $C, P, T$ and $CP$. With the use of Appendix we can see that operator structure of
these interactions are such that they acquire fermionic arbitrary phase coefficients $\eta_c^2$, $\eta_p^2$ and $\eta_t^2$ under the action of
discrete transformations, which are equal to $1$ for $\eta_c$, $\eta_p$, $\eta_t=\pm1 $ and are equal to $-1$ for $\eta_c$, $\eta_p$, $\eta_t=\pm i$.

\subsection{$\Delta B=\frac{1}{3}\Delta L=-1$ nucleon decay($p\rightarrow D^0e^{+}\bar{\nu}\bar{\nu}, p\rightarrow e^{+}e^{+}\bar{\nu}\pi^{-} etc.$) \cite{weldon1980operator}}

To construct the lowest dimensional $B-3L$ conserving $SU(3)\times SU(2)\times U(1)$ invariant operators there are two possibilities:
$l_{L}l_{L}e_{R}u_{R}u_{R}u_{R}$  and  $l_{L}l_{L}l_{L}q_{L}u_{R}u_{R}$ with $d=9$. Due to Fermi statistics $u_{R}$ fields should not be
all of the same generation. In the following operators ${u}^\prime_{R}=c_{R}$. Because of charm quark these operators can produce
energetically forbidden $p\rightarrow D^{0}e^{+}\bar{\nu}\bar{\nu}$ and can not produce $p\rightarrow e^{+}\bar{\nu}\bar{\nu}$
 \cite{weinberg1980varieties,weldon1980operator}. These operators are:

\begin{equation}\label{19}
\epsilon_{\alpha\beta\gamma}\epsilon_{ij}(l^{T}_{Li}cl^\prime_{Lj})(e_{R}^{T} cu_{R\alpha})(u_{R\beta}^{T}cu^\prime_{R\gamma})
\end{equation}
\begin{equation}
\epsilon_{\alpha\beta\gamma}\epsilon_{ij}\epsilon_{kl}(l^{T}_{Li}cl^\prime_{Lj})(l^{T}_{Lk}cq_{Ll\alpha})(u_{R\beta}^{T}cu^\prime_{R\gamma})
\end{equation}
Operators which mediate processes like $p\rightarrow e^{+}e^{+}\bar{\nu}\pi^{-}$ and $p\rightarrow e^{+}\mu^{+}\bar{\nu}\pi^{-}$
should be built with higher dimensions, $d=11$. Such operators are:
\begin{equation}\label{21}
\epsilon_{\alpha\beta\gamma}\epsilon_{ij}\epsilon_{km}\epsilon_{ln}(l^{T}_{Li}c\sigma_{\mu\nu}l_{Lj})(q^{T}_{Lk}c\sigma^{\mu\nu}q_{Ll})(u^{T}_{R}ce_{R})\phi_m \phi_n
\end{equation}

Operators \eqref{19}-\eqref{21} have the ability to annihilate three quarks in the initial state and create a net of three anti-leptons in
the final state. Therefore, they break $B-L$ symmetry and violate $F$ number. These operators have the same discrete transformations and
for the action of charge conjugation we have:
\[
  C=\left \{
  \begin{tabular}{ccc}
  terms with even number of $\gamma^{5}$ are $C-even$\\
  terms with odd number of $\gamma^{5}$ are $C-odd$
  \end{tabular}
  \right.
\]
If we change $\eta_c$ from $1$ to $\pm i$, $C-even$ terms become $C-odd$ and vice versa. For parity we have:
\[
  P=\left \{
  \begin{tabular}{ccc}
  terms with even number of $\gamma^{5}$ are $P$-odd\\
  terms with odd number of $\gamma^{5}$ are $P$-even
  \end{tabular}
  \right.
\]
If we change $\eta_p$ to $\pm i$, $P-even$ terms become $P-odd$ and vice versa.

These operators are $T-even$ and by choosing $\eta_t=\pm i$, they become $T-odd$. Also, by using our original arbitrary phase convention
these operators are $CP-odd$. Under the action of $C, P$ and $T$ these operators acquire fermionic phase coefficients $\eta_c^6$, $\eta_p^6$
and $\eta_t^6$, which are equal to $1$ for $\eta_c$, $\eta_p$, $\eta_t=\pm1 $ and are equal to $-1$ for $\eta_c$,
$\eta_p$, $\eta_t=\pm i$. Therefore, for these operators which break $B-L$ symmetry and violate $F$ number the set ($\pm 1, \pm i$)
is not the correct set of fermionic arbitrary phases, due to the fact that these values don not lead to the identical discrete transformations.

\subsection{$\Delta B=-\frac{1}{3}\Delta L=-1$ nucleon decay($n\rightarrow \nu\nu e^{-}\pi^{+}, n\rightarrow \nu\nu\nu$ etc.) \cite{weinberg1980varieties}}

The lowest dimensional $SU(3)\times SU(2)\times U(1)$ operators for $\Delta B=-\frac{1}{3}\Delta L=-1$ have d=10. The only processes of nucleon
decay which can be produced by these operators involve a single left-handed charged lepton. ($n\rightarrow \nu\nu e^{-}_L\pi^{+}, p\rightarrow \nu\nu e^{-}_L\pi^{+}\pi^+, n\rightarrow \nu\nu e^{-}_L K^{+}$)
These operators are:
\begin{equation}\label{22}
\epsilon_{\alpha\beta\gamma}\epsilon_{ij}\epsilon_{kl}(\bar{d}_{R\alpha}l_{Li})(\bar{d}_{R\beta}l_{Lj})(\bar{d}_{R\gamma}l_{Lk})\phi_{l}
\end{equation}
\begin{equation}
\epsilon_{\alpha\beta\gamma}\epsilon_{ij}\epsilon_{kl}(-d^\dagger_{R\alpha}Cd^{\star}_{R\beta})(\bar{d}_{R\gamma}l_{Li})(l^T_{Lj}Cl_{Lk})\phi_l
\end{equation}
\begin{equation}
\epsilon_{\alpha\beta\gamma}\epsilon_{il}\epsilon_{jk}(-d^\dagger_{R\alpha}Cd^{\star}_{R\beta})(\bar{d}_{R\gamma}l_{Li})(l^T_{Lj}Cl_{Lk})\phi_l
\end{equation}

To have a nucleon decays which involve right handed charged leptons or three neutrinos ($n\rightarrow\nu\nu\nu$ or $p\rightarrow\nu\nu\nu\pi^+$)
we should use operators with $d\geq 12$:
\begin{equation}\label{25}
 \epsilon_{\alpha\beta\gamma}\epsilon_{il}\epsilon_{jm}\epsilon_{kn}(\bar{u}_{R\alpha}l_{Li})(\bar{d}_{R\beta}l_{Lj})(\bar{d}_{R\gamma}l_{Lk})\phi_l\phi_m\phi_n
\end{equation}

Operators \eqref{22}-\eqref{25} are able to destroy three quarks in the initial state and create a net of three leptons in the final state.
Hence, $B-L$ symmetry is broken and $F$ number is conserved. These operators transform similarly under the action of discrete
symmetries. For charge conjugation we have:

\[
  C=\left \{
  \begin{tabular}{ccc}
  terms with even number of $\gamma^{5}$ are $C-even$\\
  terms with odd number of $\gamma^{5}$ are $C-odd$
  \end{tabular}
  \right.
\]
choosing $\eta_c= \pm i$ would not change this result. For parity we have:
\[
  P=\left \{
  \begin{tabular}{ccc}
  terms with even number of $\gamma^{5}$ are $P-even$\\
  terms with odd number of $\gamma^{5}$ are $P-odd$
  \end{tabular}
  \right.
\]
choosing $\eta_p=\pm i$ would not change this result.

The above operators are $T-even$ and changing $\eta_t$ to $\pm i$ will not change this transformation. With the use of our original phases,
these interactions are $CP-even$. As we can see, for this group of operators which break $B-L$ symmetry and conserve $F$
number choosing any value of the set ($\pm 1, \pm i$) gives identical transformations. On the other way, under the action of discrete transformations
they get the fermionic arbitrary phase coefficients $\left|\eta_c\right|^6$, $\left|\eta_p\right|^6$ and $\left|\eta_t\right|^6$ which are equal
to $1$ for any choice of arbitrary phase.

\subsection{$\Delta B=2$($n\rightarrow \bar{n}$ etc.) \cite{rao1982n}}

This process violates $B-L$ symmetry and $F$ number. A detailed study of neutron oscillation has been done by Kao and Love \cite{kuo1980neutron}
based on $SU(3)\times SU(2)\times U(1)$ gauge group and subsequently by Rao and Shrock \cite{rao1982n} based on $SU(3)\times U(1)_{em}$
gauge group and both have $d=9$. Here we consider the latter, because it requires lower energy level for this process \cite{caswell1983matter}.
But our claim about discrete transformations would be valid for operators in \cite{kuo1980neutron} as well. Neutron oscillation operators
have the form:

\begin{equation}\label{26}
(T_s)_{\alpha\beta\gamma\delta\rho\sigma}\left(u^{T\alpha}_{\chi_1}Cu^\beta_{\chi_1}\right)\left(d^{T\gamma}_{\chi_2}Cu^\delta_{\chi_2}\right)\left(d^{T\rho}_{\chi_3}Cu^\sigma_{\chi_3}\right)
\end{equation}
\begin{equation}
(T_s)_{\alpha\beta\gamma\delta\rho\sigma}(u^{T\alpha}_{\chi_1}Cd^\beta_{\chi_1})(u^{T\gamma}_{\chi_2}Cu^\delta_{\chi_2})(d^{T\rho}_{\chi_3}Cu^\sigma_{\chi_3})
\end{equation}
\begin{equation}\label{28}
(T_a)_{\alpha\beta\gamma\delta\rho\sigma}(u^{T\alpha}_{\chi_1}Cd^\beta_{\chi_1})(u^{T\gamma}_{\chi_2}Cu^\delta_{\chi_2})(d^{T\rho}_{\chi_3}Cu^\sigma_{\chi_3})
\end{equation}
where $\chi_i=R$ or $L$ and $(T_s)_{\alpha\beta\gamma\delta\rho\sigma}$ and $(T_a)_{\alpha\beta\gamma\delta\rho\sigma}$ are:
\begin{equation}\label{T_s}
(T_s)_{\alpha\beta\gamma\delta\rho\sigma}=\epsilon_{\rho\alpha\gamma}\epsilon_{\sigma\beta\delta}+\epsilon_{\sigma\alpha\gamma}\epsilon_{\rho\beta\delta}+\epsilon_{\rho\beta\gamma}\epsilon_{\sigma\alpha\delta}+\epsilon_{\sigma\beta\gamma}\epsilon_{\rho\alpha\delta}
\end{equation}
\begin{equation}\label{T_a}
(T_a)_{\alpha\beta\gamma\delta\rho\sigma}=\epsilon_{\rho\alpha\beta}\epsilon_{\sigma\gamma\delta}+\epsilon_{\sigma\alpha\beta}\epsilon_{\rho\gamma\delta}
\end{equation}
Operators \eqref{26}-\eqref{28} transform under the effect of charge conjugation in the way that:
\[
  C=\left \{
  \begin{tabular}{ccc}
  terms with even number of $\gamma^{5}$ are $C-even$\\
  terms with odd number of $\gamma^{5}$ are $C-odd$
  \end{tabular}
  \right.
\]
By choosing $\eta_c=\pm i$, $C-even$ terms become $C-odd$ and vice versa. Under the effect of parity we have:
\[
  P=\left \{
  \begin{tabular}{ccc}
  terms with even number of $\gamma^{5}$ are $P-odd$\\
  terms with odd number of $\gamma^{5}$ are $P-even$
  \end{tabular}
  \right.
\]
Due to altering $\eta_p$ to $\pm i$, $P-odd$ terms become $P-even$ and vice versa.

These operators are $T-even$ and if we change $\eta_t$ to $\pm i$ they all become $T-odd$. According to our original phase convention
operators \eqref{26}-\eqref{28} are $CP-odd$. It is clear that for this group of operators the set ($\pm 1, \pm i$)
can not be the appropriate set of fermionic arbitrary phases, because different values of this set lead to the different discrete
transformation. On the other hand, they obtain arbitrary phase coefficients $\eta_c^6$, $\eta_p^6$ and $\eta_t^6$ under the action of
discrete transformations.

\subsection{$\Delta L=2$ (neutrino masses, neutrino oscillation, etc.) \cite{weinberg1980varieties}}

The lowest dimensional $SU(3)\times SU(2)\times U(1)$ Lorentz invariant operators have d=5. These operators are:
\begin{equation}\label{31}
l^T_{Li}Cl_{Lj}\phi_k\phi_l\epsilon_{ik}\epsilon_{jl}
\end{equation}
\begin{equation}\label{32}
l^T_{Li}Cl_{Lj}\phi_k\phi_l\epsilon_{ij}\epsilon_{kl}
\end{equation}

Interactions which can be produce by operators \eqref{31}-\eqref{32}, such as neutrino oscillation or neutrinoless double$-\beta$ decay
($nn\rightarrow e^{-}e^-pp$ by Eq.\eqref{32}), violate $B-L$ symmetry and $F$ number. Beside the above operators, higher dimension
$\Delta L=2$ operators with $d=7$ has been considered in reference \cite{weldon1980operator}. Here we are interested in the transformation of
these operators under the effect of discrete symmetries and all of these operators transform in the same way as follows. Under the action of
charge conjugation we have:
\[
  C=\left \{
  \begin{tabular}{ccc}
  terms with even number of $\gamma^{5}$ are $C-even$\\
  terms with odd number of $\gamma^{5}$ are $C-odd$
  \end{tabular}
  \right.
\]
if we change the arbitrary phase $\eta_c$ to $\pm i$ then $C-even$ terms become $C-odd$ and vice versa. Under the action of parity
we have:
\[
  P=\left \{
  \begin{tabular}{ccc}
  terms with even number of $\gamma^{5}$ are $P-odd$\\
  terms with odd number of $\gamma^{5}$ are $P-even$
  \end{tabular}
  \right.
\]
if we change $\eta_p$ to $\pm i$ then $P-odd$ terms become $P-even$ and vice versa.

Both of these operators are $T-even$ and by choosing $\eta_t=\pm i$ they become $T-odd$, and based on our original phase convention they
are  $CP-odd$. Thus, in the case of these operators which break $B-L$ and violate $F$ number, the set ($\pm 1, \pm i$)
is not the correct set of fermionic arbitrary phase. On the other hand, under the action of discrete transformation they get fermionic
phase coefficients $\eta_c^2$, $\eta_p^2$ and $\eta_t^2$.

\subsection{ $\Delta B=\Delta L=-2$($H\bar{H}$ transition, etc.) \cite{caswell1983matter}}
The $d=12$ operators which mediate $H\bar{H}$ transition and double proton decay ($pp\rightarrow e^+e^+$) conserve $B-L$ symmetry and
break $F$ number:
\begin{equation}\label{33}
(e^T_{\chi_1}Ce_{\chi_1})(u^{T\alpha}_{\chi_2}Cu^\beta_{\chi_2})(d^{T\gamma}_{\chi_3}Cu^\delta_{\chi_3})(d^{T\rho}_{\chi_4}Cu^\sigma_{\chi_4})(T_s)_{\alpha\beta\gamma\delta\rho\sigma}
\end{equation}
\begin{equation}
(u^{T\alpha}_{\chi_1}Cu^\beta_{\chi_1})(u^{T\gamma}_{\chi_2}Cu^\delta_{\chi_2})(d^{T\rho}_{\chi_3}Ce_{\chi_3})(d^{T\sigma}_{\chi_4}Ce_{\chi_4})(T_s)_{\alpha\beta\gamma\delta\rho\sigma}
\end{equation}
\begin{equation}
(u^{T\alpha}_{\chi_1}Cu^\beta_{\chi_1})(d^{T\gamma}_{\chi_2}Cd^\delta_{\chi_2})(u^{T\rho}_{\chi_3}Ce_{\chi_3})(u^{T\sigma}_{\chi_4}Ce_{\chi_4})(T_s)_{\alpha\beta\gamma\delta\rho\sigma}
\end{equation}
\begin{equation}
(u^{T\alpha}_{\chi_1}Cu^\beta_{\chi_1})(u^{T\gamma}_{\chi_2}Cd^\delta_{\chi_2})(u^{T\rho}_{\chi_3}Ce_{\chi_3})(d^{T\sigma}_{\chi_4}Ce_{\chi_4})\left\{\begin{matrix}
(T_s)_{\rho\sigma\gamma\delta\alpha\beta}
\\ (T_a)_{\rho\sigma\gamma\delta\alpha\beta}
\end{matrix}\right.
\end{equation}
\begin{multline}\label{37}
 (u^{T\alpha}_{\chi_1}Cd^\beta_{\chi_1})(u^{T\gamma}_{\chi_2}Cd^\delta_{\chi_2})(u^{T\rho}_{\chi_3}Ce_{\chi_3})(u^{T\sigma}_{\chi_4}Ce_{\chi_4}) \\
 \begin{Bmatrix} {(T_s)_{\alpha\beta\gamma\delta\rho\sigma},(T_a)_{\alpha\beta\gamma\delta\rho\sigma},(T_a)_{\rho\sigma\gamma\delta\alpha\beta},(\tilde{T}_a)_{\alpha\beta\gamma\delta\rho\sigma}} \end{Bmatrix}
\end{multline}
where $T_s$ and $T_a$ are the same as eqs. \eqref{T_s} and \eqref{T_a} and $\tilde{T}_a$ is:
\begin{equation}
(\tilde{T}_a)_{\alpha\beta\gamma\delta\rho\sigma} = \epsilon_{\alpha\beta\rho} \epsilon_{\gamma\delta\sigma} - \epsilon_{i\beta\sigma} \epsilon_{\gamma\delta\rho}
\end{equation}

For the transformation of these operators under charge conjugation we have:
\[
  C=\left \{
  \begin{tabular}{ccc}
  terms with even number of $\gamma^{5}$ are $C$-even\\
  terms with odd number of $\gamma^{5}$ are $C$-odd
  \end{tabular}
  \right.
\]
changing $\eta_c$ to $\pm i$ would not alter the above transformations. For parity we have:
\[
  P=\left \{
  \begin{tabular}{ccc}
  terms with even number of $\gamma^{5}$ are $P$-even\\
  terms with odd number of $\gamma^{5}$ are $P$-odd
  \end{tabular}
  \right.
\]
choosing $\eta_p=\pm i$ would not change the above transformations.

All of the above operators are $T-even$ and by choosing $\eta_t=\pm i$ this results would not change. These operators are $CP-even$
as well. As we can see for operators \eqref{33}-\eqref{37} which conserve $B-L$ and break $F$ number discrete
transformations will not change by switching fermionic arbitrary phases from $\pm 1$ to $\pm i$, due to the fact that they acquire
phase coefficients $\eta_c^8$, $\eta_p^8$ and $\eta_t^8$.

\section{Conclusion}

In this paper we have collected baryon- and lepton-nonconserving operators from different references to evaluate discrete transformations
of them.

It has long been assumed that the values ($\pm1, \pm i$) form the correct set of arbitrary phases for fermionic field operators.
In this paper we have shown that for any operator which break $B-L$ symmetry and violate $F$ number the set
($\pm 1, \pm i$) is not the correct set of fermionic arbitrary phases. For this class of operators changing the phases form $\pm 1$ to
$\pm i$ will not lead to the same discrete transformations. Therefor, the set ($\pm 1, \pm i$)
do not form the appropriate set of arbitrary phases for this class of operators.

\section{ACKNOWLEDGMENTS}
I am grateful for valuable discussion with  Masud Chaichian and Michael Eides. When we were working on the last version of this
manuscript, we came across a paper by G. Feinberg and S. Weinberg \cite{feinberg1959phase} in which they have considered parity
transformation of Hamiltonians which need different set of fermionic arbitrary phases from the set ($\pm 1, \pm i$).

\section{Appendix}

In the following table we list discrete transformation of fermion bilinears under the action of $C$, $P$ and $T$. These fermion bilinears
are used as building blocks of the interacting operators. We do not choose any specific value for the arbitrary phases and they are  written
as $\eta_c$, $\eta_p$ and $\eta_t$. It is easy to see that under the action of discrete transformations, fermion bilinears in the form of
$\bar{\psi}(\gamma-matrices)\psi$ acquire arbitrary phase coefficients $ \left|\eta_c\right|^2$, $\left|\eta_p\right|^2$ and
$\left|\eta_t\right|^2$ which is equal to $1$ for any value of arbitrary phases. But fermion bilinears in the form of
$\psi^T(\gamma-matrices) \psi$ acquire arbitrary phase coefficients $\eta_c^2$, $\eta_p^2$ and $\eta_t^2$ which are equal to $1$ for
$\eta_c=\eta_p=\eta_t=\pm1$ and they are equal to $-1$ for $\eta_c=\eta_p=\eta_t=\pm i$.

By the shorthand notation $(-1)^\mu$ we mean, $(-1)^\mu\equiv1$ for $\mu=0$ and $(-1)^\mu\equiv-1$ for $\mu=1,2,3.$

\begin{center}
\renewcommand{\arraystretch}{1.3}
\begin{tabular}{ c|c|c|c  }
   & C & P & T  \\
  \hline
  $\psi^Tc\psi$ & $+\eta_c^2$ & $-\eta_p^2$ & $+\eta_t^2$  \\
  $\psi^Tc\gamma^5\psi$ & $-\eta_c^2$ & $+\eta_p^2$ & $+\eta_t^2$  \\
  $\bar{\psi}\psi$ & $+1$ & $+1$ & $+1$ \\
  $\bar{\psi}\gamma^5 \psi$ & $-1$ & $-1$ & $+1$ \\
  $\psi^TcD^\mu\psi$ & $-\eta_c^2$ & $-(-1)^\mu\eta_p^2$ & $-(-1)^\mu\eta_t^2$  \\
  $\psi^TcD^\mu\gamma^5\psi$ & $+\eta_c^2$ & $(-1)^\mu\eta_p^2$ & $-(-1)^\mu\eta_t^2$  \\
  $\bar{\psi}\gamma^{\mu}\psi$ & $-1$ & $(-1)^\mu$ & $(-1)^{\mu}$ \\
  $\bar{\psi}\gamma^\mu \gamma^5 \psi$ & $+1$ & $-(-1)^\mu$ & $(-1)^\mu $ \\
  $\psi^Tc\gamma^\mu\psi$ & $-\eta_c^2$ & $-(-1)^\mu\eta_p^2$ & $(-1)^\mu\eta_t^2$ \\
  $\psi^Tc\gamma^\mu\gamma^5\psi$ & $+\eta_c^2$ & $(-1)^\mu\eta_p^2$ & $(-1)^\mu\eta_t^2$ \\
  $\bar{\psi}D^\mu \psi$ & $+1$ & $(-1)^\mu$ & $-(-1)^\mu1$ \\
  $\bar{\psi}D^\mu \gamma^5 \psi$ & $-1$ & $-(-1)^\mu$ & $-(-1)^\mu$ \\
  $\psi^Tc \sigma^{\mu\nu}\psi$ & $-\eta_c^2$ & $-(-1)^\mu(-1)^\nu\eta_p^2$ & $-(-1)^\mu(-1)^\nu\eta_t^2$ \\
  $\psi^Tc \sigma^{\mu\nu}\gamma^5\psi$ & $+\eta_c^2$ & $(-1)^\mu(-1)^\nu\eta_p^2$& $-(-1)^\mu(-1)^\nu\eta_t^2$
  \end{tabular}
\end{center}


\end{document}